\documentclass[nofootinbib,amsmath,amssymb,aps,prl,reprint,noeprint,floatfix,superscriptaddress]{revtex4-2}

\usepackage{float}
\usepackage{graphicx}
\usepackage{dcolumn}
\usepackage{bm}
\usepackage[version=4]{mhchem}
\usepackage[breaklinks,colorlinks,linkcolor={blue}, %
            citecolor={blue},urlcolor={blue}]{hyperref}

\usepackage{physics}
\usepackage{amssymb}
\usepackage{amsfonts}         
\usepackage{graphicx}         
\usepackage[T1]{fontenc}     
\usepackage{ae}               
\usepackage{lastpage}         
\usepackage{textcomp}
\usepackage{tabularx}
\usepackage{xcolor}
\usepackage{color}
\usepackage{nicefrac}
\usepackage{upgreek}
\usepackage{subcaption} 
\usepackage{float}
\usepackage{mathtools}
\usepackage{gensymb}

\begin{document}

\title{Broad-Range Directional Detection of Light Dark Matter in Cryogenic Ice} 

\author{Nora Taufertsh{\"o}fer}
\affiliation{Molecular Foundry Division, Lawrence Berkeley National Laboratory, Berkeley, California 94720, USA}
\affiliation{Materials Sciences Division, Lawrence Berkeley National Laboratory, Berkeley, California 94720, USA}
\affiliation{Institute for Theoretical Physics, University of W\"urzburg, Am Hubland, D-97074 W\"urzburg, Germany}
\author{Maurice Garcia-Sciveres}
\affiliation{Physics Division, Lawrence Berkeley National Laboratory, Berkeley, California 94720, USA}
\author{Sin{\'e}ad M. Griffin}
\email{sgriffin@lbl.gov}
\affiliation{Molecular Foundry Division, Lawrence Berkeley National Laboratory, Berkeley, California 94720, USA}
\affiliation{Materials Sciences Division, Lawrence Berkeley National Laboratory, Berkeley, California 94720, USA}

\date{\today}

\begin{abstract}

We propose hexagonal ice (H$_2$O) as a new target for light dark matter (DM) direct detection. Ice, a polar material, is suitable for single phonon detection through DM scattering for which we consider light dark photon and light scalar mediator models. We report a rate sensitivity down to a DM mass of $\sim$\,keV, constituting a broader mass range than other promising candidates. We find better sensitivity for near-term experimental thresholds from the presence of high-frequency phonons. These advantages, and ice's availability, make it highly promising for single-phonon detection.
\end{abstract}

\maketitle

New proposals for the direct detection of dark matter (DM) have steadily pushed into the sub-GeV `light' DM range\cite{Essig_et_al:2012}. In particular, for DM masses in the kev-GeV range, freeze-in DM scattering is well motivated through both a light scalar mediator and a dark photon mediator\cite{Hall_et_al:2010,Bernal_et_al:2017}. The smaller kinetic energy for this lower mass DM motivates the exploration of low-energy excitations in condensed matter systems as possible direct detection channels\cite{Kahn/Lin:2022}. 
Recent proposals for DM detection in crystal targets include quasiparticle excitations ranging from phonon and electron scattering in conventional semiconductors\cite{Essig_et_al:2016,Knapen_2018_polar_materials} to more exotic quantum materials such as Dirac semimetals\cite{Hochberg_et_al:2018}, and axions in topological insulators and multiferroics\cite{Marsh_et_al:2019,Roising_et_al:2021, Engel_et_al:2021}.  

Of these proposals for DM direct detection via quasiparticle scattering, single phonon production presents a promising detection scheme for several reasons\cite{Knapen_2018_polar_materials,Griffin_GaAs_Sapphire}. Typical optical phonons in crystals range from 10-100 meV, making them kinematically matched with sub-MeV DM. Dark photon mediators can mix with standard model photons resulting in a coupling to optical phonon modes in polar semiconductors. Furthermore, common polar semiconductors (e.g. GaAs) have sizeable bandgaps resulting in small screening. The intrinsic anisotropy of polar semiconductors allows for a directional response giving signal discrimination based on daily/annual modulation of the DM wind\cite{Drukier_et_al:1986}. Importantly, high-quality detector-grade crystals of a range of polar semiconductors are readily available owing to their use in microelectronics and quantum computing  (e.g. GaAs\cite{Knapen_2018_polar_materials}, SiC\cite{Griffin_et_al:2021}), and phonon sensing technologies are steadily improving towards the low sensitivities required for detection\cite{Pyle_et_al:2015}.

With the next-generation of low-mass DM experiments in planning, it remains to identify the best possible target material for each detection scheme. Previous work addressed this at the DM-target interaction level by performing a comparison of DM-phonon reach, isolating materials-specific factors that optimize for low-mass sensitivity and greatest cross section\cite{Griffin_et_al:2020}. Beyond this, other materials factors for target selection should also be considered that take into account current synthesis/fabrication capabilities and sensing constraints for near-term experiments. While experimental thresholds for phonon sensing continue to improve towards the single phonon excitation regime ($\sim$10-100 meV), near-term experiments with larger thresholds could access higher energy responses. For instance, beyond the single phonon excitation regime, multiphonon responses can be accessed for DM scattering above the single phonon spectra though at a suppressed rate\cite{Campbell_et_al:2020}. However, an alternative approach to this is to search for single phonon and other quasiparticle responses that lie within the current sensing thresholds (>100 meV). 

In this \textit{Letter}, we propose cryogenic water ice (H$_2$O) as a particularly appealing target material for direct detection of low-mass DM from single-phonon excitations. We find that the mixed nature of bonding in H$_2$O results in a broad range of phonon frequencies suitable for both lower-mass detection, and for higher frequency single phonon excitations that are within the range of current sensing thresholds. Moreover, with the ubiquity and long history of fabrication and characterization of solid H$_2$O, it can be readily fabricated as large single crystals comprising earth-abundant (inexpensive) elements.

This paper is laid out as follows; we summarize the calculation of spin-independent DM scattering for single-phonons and the features of ice H$_2$O (specifically its XI$_h$ polymorph) of relevance to DM-phonon interactions. We then present our calculations of the reach for a light dark photon mediator and a light scalar mediator with comparisons to best-performing targets for these DM models, and their calculated directional dependence. Finally we discuss other material limiting considerations for target materials and prospects for its use for near-term and next-generation DM detection. 

First, we briefly summarize the formalism for calculating spin-independent DM scattering processes including a light dark photon mediator and a light scalar mediator\cite{TheoFramework}.
For a detector with target density $\rho_{\text{T}}$ the total rate per target mass is given by 

\begin{align}
    R =\frac{1}{\rho_{\text{T}}} \frac{\rho_{\chi}}{m_{\chi}} \int d^3v f_{\chi}(\mathbf{v}) \Gamma(\mathbf{v}) 
    \end{align}

where $\rho_{\chi}$ is the local DM energy density, $m_{\chi}$ the DM mass, and $f_{\chi}(\mathbf{v})$ is the DM velocity distribution in the lab frame which is modelled with a boosted Maxwell-Boltzmann distribution \footnote{Due to Earth's rotation the orientation of the incoming DM particles with regard to the target crystal changes over a day, which leads to a directional modulation of the rate. We choose the z-axis in the crystal frame to be parallel to Earth's velocity at daytime $t=0$\cite{Lewin/Smith:1996}.}. $\Gamma(\mathbf{v})$ is the scattering rate per non-relativistic DM particle which is defined by an integral over the momentum transfer $\mathbf{q}=\mathbf{p}-\mathbf{p}'$ with initial DM momentum $\mathbf{p}=m_{\chi} \mathbf{v}$:
    \begin{align}
       \Gamma(\mathbf{v}) 
       = \frac{\pi \bar{\sigma}}{\mu^2} \int \frac{d^3q}{(2\pi)^3} (q_0/q)^4 S(\mathbf{q}, \omega_{\mathbf{q}})
    \end{align}
%
Here, $\bar{\sigma}$ is a model-dependent reference cross section and we define  $\bar{\sigma} := \frac{\mu^2}{\pi} \overline{|\mathcal{M}(q_0)|^2}$ with $\mathcal{M}$ the target-independent $2 \rightarrow 2$ scattering matrix element (see Supplemental Material). 
$\mu$ is the reduced mass of an electron or nucleon and DM particle for light-dark-photon- or light-scalar-mediated scattering, respectively. $q_0$ is a reference momentum and $S(\mathbf{q},\omega_{\mathbf{q}})$ the target-dependent dynamic structure factor. 
For single phonon excitations we recall the dynamical structure factor:
%
    \begin{align}
    \begin{split}
        &S(\mathbf{q}, \omega_{\mathbf{q}}) = \\
        &\frac{\pi}{\Omega} \sum_{\nu} \frac{1}{\omega_{\nu, \mathbf{k}}}
        \Big  \vert \sum_j \frac{e^{-W_j(\mathbf{q})}}{\sqrt{m_j}} \,
        e^{i \mathbf{G}\cdot \mathbf{x}^0_j}\,
        (\mathbf{Y}_j \cdot \boldsymbol{\epsilon}^{\ast}_{\nu, j, \mathbf{k}}) \Big \vert ^2 
        \delta (\omega_{\mathbf{q}} - \omega_{\nu, \mathbf{k}})
    \end{split}
    \end{align}
where $\omega_{\mathbf{q}} = \mathbf{q} \cdot \mathbf{v} -\frac{q^2}{2m_{\chi}}$ explicitly depends on the energy deposition. The sums run over all phonon branches $\nu$ and over all ions $j$ in the primitive cell. The ionic masses are $m_j$ with equilibrium positions $\mathbf{x}^0_j$ and $\Omega$ is the primitive cell volume. $\mathbf{\epsilon}_{\nu, j, \mathbf{k}}$ defines the phonon polarization vectors. $\mathbf{G}$ is the reciprocal lattice vector that satisfies $\mathbf{G} = \mathbf{q} - \mathbf{k}$ with $\mathbf{k}$ within the first Brillouin zone. DM couplings come in with the model-specific $\mathbf{Y}_j$ terms defined below. In the continuum limit for $\mathbf{k}$, the Debye-Waller factor is given by:
%
    \begin{align}
        W_j(\mathbf{q}) =
        \frac{\Omega}{4 m_j} \sum_{\nu} \int_{1\text{BZ}}
        \frac{d^3k}{(2 \pi)^3} 
        \frac{|\mathbf{q} \cdot \boldsymbol{\epsilon}_{\nu, j, \mathbf{k}}|^2}{\omega_{\nu, \mathbf{k}}}.
    \end{align}
The materials-dependent quantities, namely the phonon frequencies $\omega_{\nu, \mathbf{k}}$ for branch $\nu$ and momentum $\mathbf{k}$, and the phonon polarization vectors $\mathbf{\epsilon}_{\nu, j, \mathbf{k}}$, can be calculated using first-principles methods such as Density Functional Theory\cite{George_et_al:2020}.

Finally, we consider two well-motivated models of light DM, namely a kinematically mixed dark photon and a light scalar mediator\cite{Knapen_2018_polar_materials}. In the former case, the standard model photon kinematically mixes with the dark photon owing to their same quantum numbers, resulting in `millicharged' DM under $U(1)$ of the standard model. In the $q \rightarrow 0$ limit valid for light DM scattering\cite{TheoFramework}, $\mathbf{Y}_j$ is given by
    \begin{align}
    \label{Y_dark_photon}
       \mathbf{Y}_{j, \text{dark photon}} = - \frac{q^2}{\mathbf{q} \cdot \varepsilon_{\infty} \cdot \mathbf{q}} (\mathbf{q} \cdot \mathbf{Z}^{\ast}_j)
    \end{align}
where $\varepsilon_{\infty}$ is the high-frequency dielectric tensor and $\mathbf{Z}^{\ast}_j$ is the Born effective charge tensor of the $j$th atom in the unit cell, both of which can be calculated using first-principles methods.
%
%
We also consider DM that only couples to nucleons via a light scalar mediator with identical coupling for proton and neutron.
The now scalar-valued $\mathbf{Y}_j$ is given by
    \begin{align}
    \label{Y_hadrophil_scalar_med}
       \mathbf{Y}_{j,\text{hadrophilic scalar}} = A_j F_{N_j}(q)
    \end{align}
%
with $A_j$ the atomic mass number and $F_{N_j}$ an isotropic nuclear form factor of the $j$th atom, respectively. The derivation of (\ref{Y_hadrophil_scalar_med}) is valid for the $q \rightarrow 0$ limit, which is fulfilled for the single phonon excitation regime. Further, for this type of scattering $F_{N_j}$ can be set to one.

\begin{figure*}[t]
\begin{center}
\includegraphics[trim = 10mm 55mm 20mm 60mm, clip, width=1\linewidth]{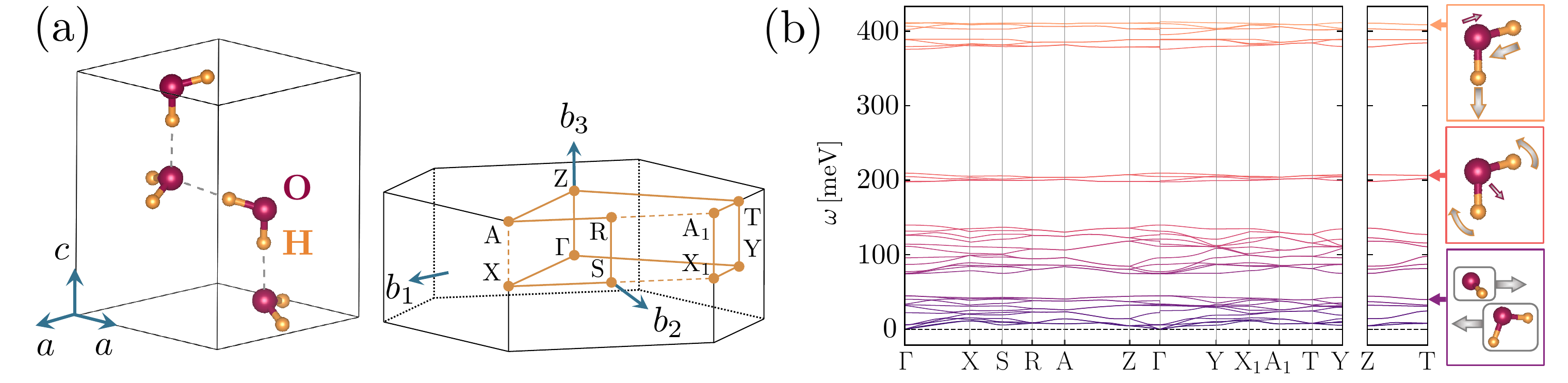}
\caption[]{(a) Primitive unit cell of hexagonal ice XI$_h$ and its first Brillouin zone with high-symmetry points labelled. (b) Calculated phonon band structure for hexagonal ice XI$_h$. The cartoons on the right hand side show the movement of ions for three selected phonon modes representative of the types of optical phonon modes in ice: translational modes (bottom), libration/bending (middle), and stretch (top).}
\label{XIh_Crystal_1BZ_Phonons}
\end{center}
\end{figure*}
The pervasiveness of liquid and solid H$_2$O has generated extensive studies of the phase diagram of ice. So far, twenty polymorphs of crystalline H$_2$O have been identified under various conditions of temperature and pressure, with the most recent, ice XIX reported in 2021\cite{Hansen_hunt_icephases}. Common to all of these is the local atomic-scale arrangement of hydrogen atoms in bonded units that fulfil the `Bernal-Fowler ice rules': each oxygen forms a covalent bond with two hydrogens and a weaker van-der-Waals bond with two other hydrogens\cite{Bernal_Fowler_ice_rules}. Since each oxygen is also shared between two hydrogen atoms, this results in an average oxygen bonding environment comprising `two-in' (strong) and `two-out' (weak) bonds with hydrogen. Such an atomic arrangement is geometrically frustrated on a tetrahedral lattice\cite{Griffin/Spaldin:2017}, resulting in disordered relative arrangements of the strong and weak bonding networks known as proton disorder. Common ice (I$_h$) forms such a hydrogen-disordered network of H-O-H bonds at ambient pressure, giving rise to a residual entropy as first predicted by Pauling\cite{Pauling:1935}. 

However, spontaneous ordering of the hydrogen networks can occur on cooling, or by the introduction of a dopant to overcome the kinetic barrier to ordering. For example KOH doping of I$_h$ results in the formation of a fully ordered phase of hexagonal ice, XI$_h$ below $\sim$ 72 K\cite{transition_72K_XIh,KOH_doping_1982}. In fact, the hydrogen ordering in the XI$_h$ structure gives it a net dipole moment, making it potentially ferroelectric\cite{ice_XI_not_that_ferroelectric}. Since XI$_h$ is the stable form of ice at ambient pressures, and can be synthesized in its ordered form, we focus on this polymorph for the remaining discussion, noting, however, that other structures of ice should have comparable DM reach for single phonon-based detection owing to their similar bonding networks. We also considered heavy ice D$_2$O XI$_h$ but we did not find a significant difference to H$_2$O (results reported in the Supplemental Material).

We now present our Density Functional Theory (DFT) calculations of the structural and phonon properties of ice XI$_h$. The full calculation details can be found in the Supplemental Material. With its combination of covalent bonding, hydrogen bonding, and van-der-Waals bonding, the challenge of treating ice H$_2$O with DFT functionals has been explored extensively\cite{Santra_et_al:2013}. We benchmark our choice of exchange correlation functional for accurate structural and vibrational properties. Consistent with previous work\cite{Raza_ferroelectric_Ic_moststable}, we find that the van-der-Waals corrected nonlocal functional (optPBE-vdw) of Klime\v{s} et al. performs best\cite{optPBE_vdW-DF_1, optPBE_vdW-DF_2}, resulting in the lattice constants $a=4.470$\,Å and $c=7.212$\,Å. This corresponds to 0.6$\%$ and 1.5$\%$ deviation from experimental measurements at T = 2\,K\cite{Fortes2018}.

In Fig.~\ref{XIh_Crystal_1BZ_Phonons}(b) we plot the calculated phonon dispersion for ice XI$_h$. All polymorphs of ice exhibit a large range of phonon frequencies as a result of the varied nature of bonding in H$_2$O crystals\cite{anom_isotropic_properties_ice_phonon_lifetimes, lattice_vib_ice_Ic,Hydrogen_bond_vib_XII}. The low-energy optical phonon range is made up of `translational' modes where individual H$_2$O molecules behave as atom-like clusters, and vibrate with respect to one another. Such low-frequency phonon modes are common in molecular crystals such as H$_2$O where molecular units are weakly bonded to each other to form the crystal network\cite{Brown_et_al:2016}. The next highest energy set of phonon modes corresponds to libration and bending of H-O bonds, with the highest-frequency range caused by the stretching of the O-H bond. As the lightest element in the periodic table, hydrogen sets an upper limit on the range of these high-frequency stretch modes that can be found in a material.

\begin{figure}[htbp]
    \begin{center}
\includegraphics[trim = 15mm 45mm 180mm 55mm, clip, width=1\linewidth]{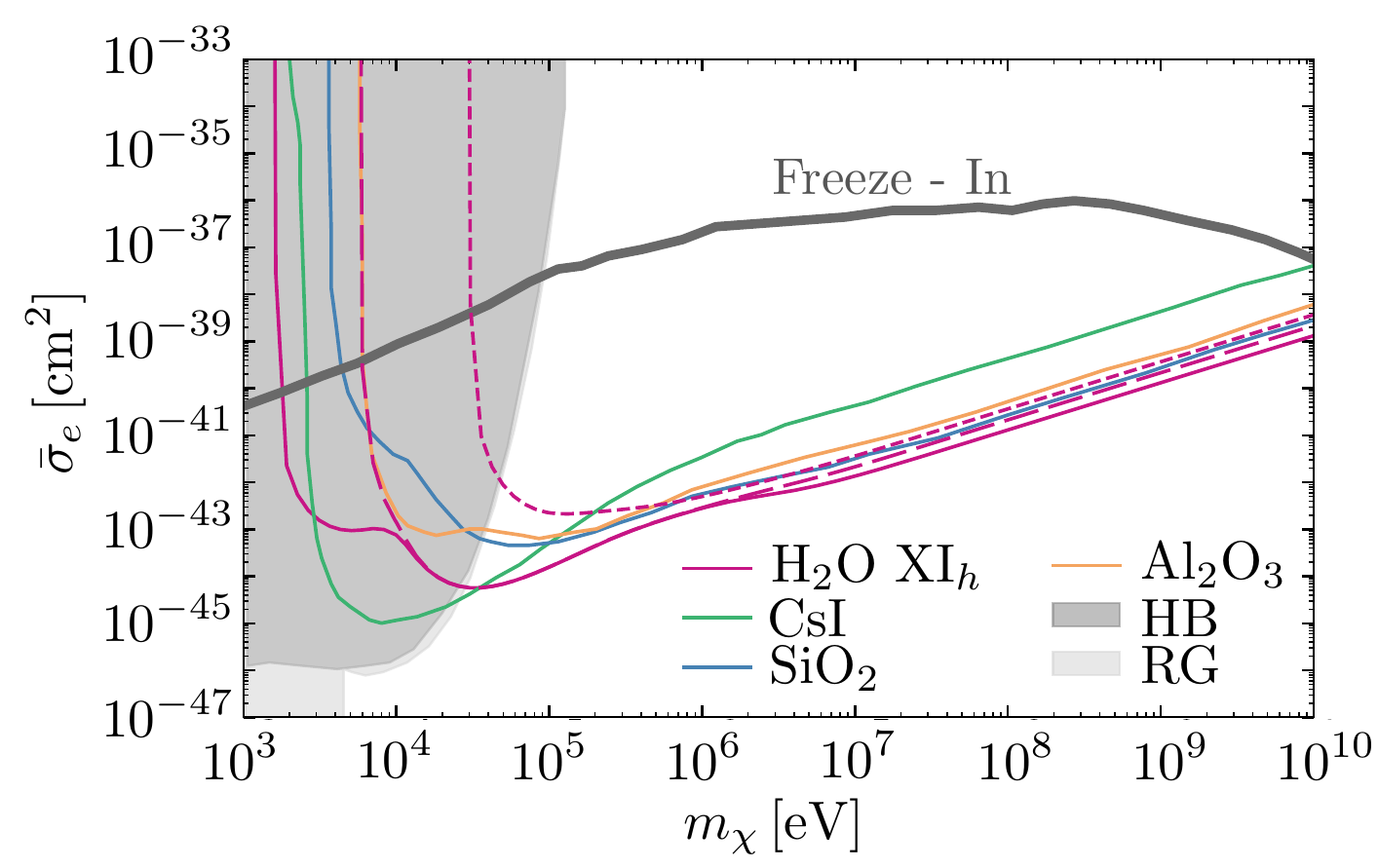}
    \caption[]{Projected reach for light-dark-photon-mediated DM scattering via single phonon excitations with 1kg-year exposure and no background. Solid,  long-dashed and dashed lines refer to a  1\,meV, 20\,meV and 100\,meV detector threshold. For comparison, additional reach data \cite{Griffin_et_al:2020} is shown. Stellar constraints from red giants (RG) and the horizontal branch (HB) are taken from \cite{stellar_constraints_dark_photon}, the freeze-in as referenced in \cite{Griffin_et_al:2020}.}
    \label{reach_comparison_darkphoton}
    \end{center}
\end{figure}

We present our calculated reach for light-dark-photon-mediated scattering of single phonons in ice XI$_h$ in Fig.~\ref{reach_comparison_darkphoton}, including comparisons to top-performing targets previously studied\cite{Griffin_et_al:2020}. All DM reach calculations were performed with the code PhonoDark\cite{Tanner_Eff_Field_theory_PhonoDark, Phonodarkcode} and our own DFT calculations. The reach of ice XI$_h$ extends further into both the low-mass DM range (well into the constrained regions) and has better reach in the high-mass DM range than any other previously suggested single target material. To understand the exceptionally broadband sensitivity of ice XI$_h$, we evaluate its performance with respect to our previously suggested quality factors for dark photon mediators in the high- and low-frequency phonon ranges\cite{Griffin_et_al:2020}. The lowest DM mass accessible is determined by the energy of the lowest-frequency optical mode where $m_\chi \sim \frac{1}{3}\omega_0^{min}\times10^6$, suggesting that materials with low-lying optical phonons such as CsI ($\omega_0^{min} \sim 7$ meV) have the best low-mass reach. In our case, the van-der-Waals bonded molecular units also have very low lying optical modes ($\omega_0^{min} \sim 6$ meV), making ice XI$_h$ optimal for the low-mass DM range. We expect all similar molecular crystals with such low-frequency translational phonon modes to have competitive reach for low-mass DM. 

For the high-mass range, a quality factor, $Q$, is defined as\cite{Griffin_et_al:2020}:
\begin{equation}
    Q \equiv \frac{Z^{\ast2}}{A_{1}A_{2}\varepsilon_{\infty}^{2}} \left(\frac{\text{meV}}{\omega_{LO}}\right)
\end{equation}
where $Z^{\ast}$ are the Born effective charges, $A_1$, $A_2$ are the atomic masses, $\varepsilon_\infty$ is the high-frequency dielectric constant, and $\omega_{LO}$ is the longitudinal optical phonon frequency. A high $Q$ corresponds to better reach in the high-mass DM regime. We calculate each of these values for ice XI$_h$ using DFT, with the full set of results given in the Supplemental Material. The best performing materials previously proposed include LiF with $Q = 270\times 10^{-7}$ and SiO$_2$ with $Q = 200\times 10^{-7}$. For ice XI$_h$, we find $Q = 533\times 10^{-7}$ for its highest-frequency stretch modes (>375 meV), steadily increasing for lower-frequency $\omega_{LO}$ clusters. The substantially enhanced $Q$ of ice XI$_h$ in the high-mass range is due to the low masses of hydrogen and oxygen. This optimized reach is dominated by hydrogen's extremely low mass, despite these small masses also resulting in higher frequency phonon modes. However, this quality factor does not take the detection threshold into account which is especially important for near-term experiments. 
%
%
We conclude that for optimal high-mass DM single-phonon-based detection with dark photon mediators, chemistries that include light elements such as H and He, and especially H-H bonds, will give the best reach in the high-mass range owing to their high-frequency single phonon modes. In the low-mass range, we find that both compounds containing heavy elements and molecular crystals that possess low-frequency optical phonon modes will give the best sensitivity. With ice combining both of these properties (molecular units and H-H bonds), it displays better sensitivity for dark photon mediators than previously suggested targets across a broad range of DM masses with CsI being the only exception for a very small low-mass range (Fig.~\ref{reach_comparison_darkphoton}).

\begin{figure}[t]
    \begin{center}
    \includegraphics[trim = 15mm 45mm 180mm 55mm, clip, width=1\linewidth]{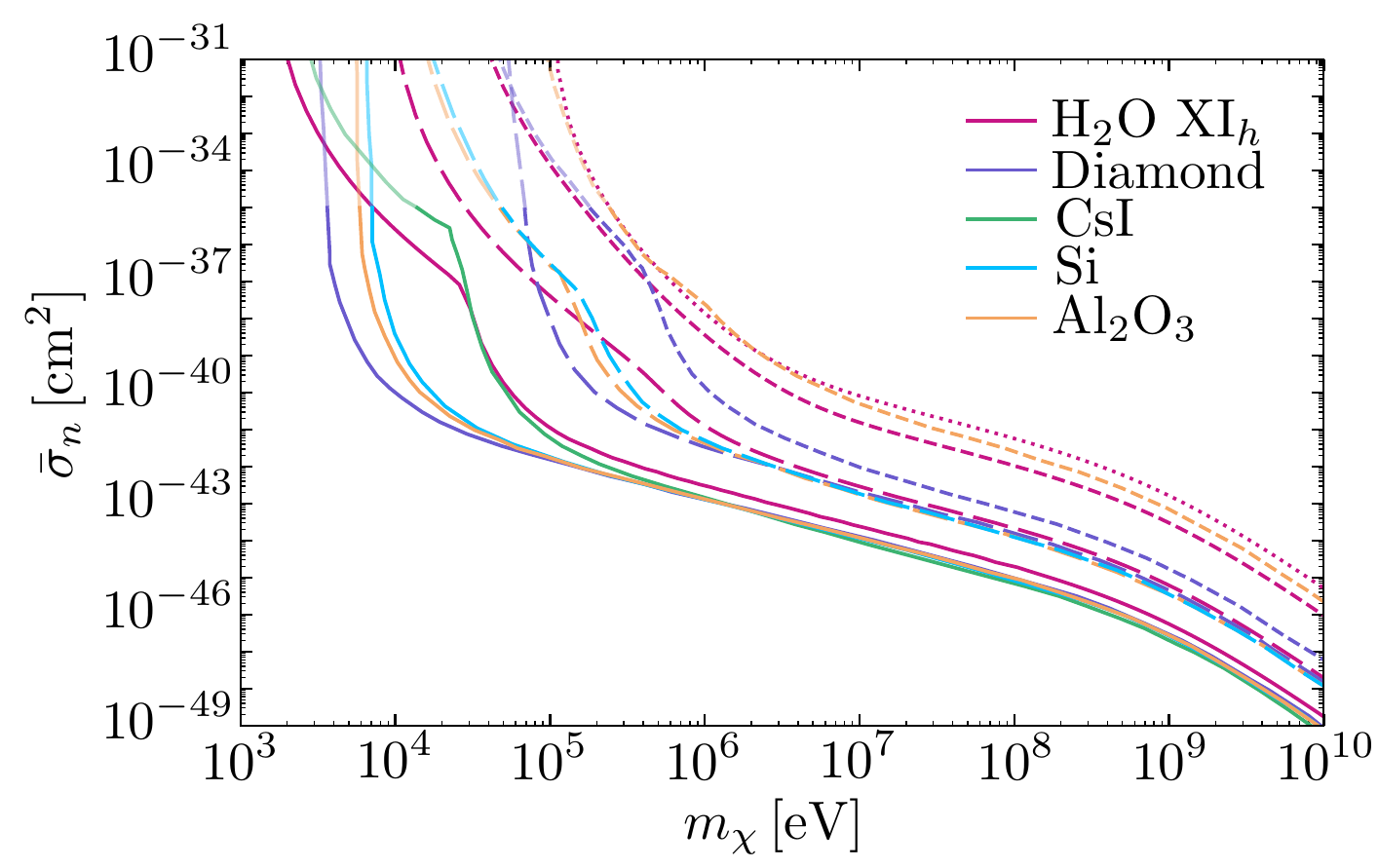}
    \caption[]{Projected reach for light-scalar-mediated DM scattering via single phonon excitations with 1kg-year exposure and no background. Solid,  long-dashed, dashed and dotted lines refer to a  1\,meV, 20\,meV, 100\,meV and 400\,meV detector threshold. Additional reach data is taken from \cite{Griffin_et_al:2020}, light-colored lines show the expected curve progression for scale ranges not shown in the reference data. }
    \label{reach_comparison_lightscalar}
    \end{center}
\end{figure}

The reach for light-hadrophilic-scalar-mediated scattering of single phonons in ice XI$_h$ is plotted in Fig.~\ref{reach_comparison_lightscalar} for several detector thresholds, and compared to previous top candidates from Ref.~\cite{Griffin_et_al:2020}. 
For the light scalar mediator, we also find that ice XI$_h$ has broad-band sensitivity for both low- and high-mass DM. Looking at the reach curves, we find a competition between which of diamond or ice is favorable for different mass ranges and thresholds. For all thresholds considered, we find that ice has sensitivity to lower masses than diamond, which become comparable by a threshold of 100 meV. In the low mass range, the lowest DM mass sensitivity is generally governed by $\omega_{min}$/$c_s$ where $\omega_{min}$ is the detector threshold and $c_s$ is the material's speed of sound. Therefore, materials with higher $c_s$ such as diamond will have the best sensitivity to lower-mass DM given low detector thresholds. We calculated the directional averaged speed of sound of ice XI$_h$ to be 4376 ms$^{-1}$ (details given in the Supplemental Material), which is a factor of three smaller than in diamond. However, for materials with smaller $c_s$ the acoustic phonons can be kinematically inaccessible, especially as the threshold increases to higher energies. When this happens, the lowest DM sensitivity is then set by the lowest optical phonon available. For ice XI$_h$ we find that the minimum reachable DM mass from kinematic considerations is $\sim$ 29 keV for a 1 meV threshold, while the lowest optical phonon is 6 meV. This manifests as a change in the slope as seen in Fig.~\ref{reach_comparison_lightscalar} where optical phonons, rather than acoustic phonons, dominate the low mass response.

In the high-mass, high-threshold range (detector threshold of 400 meV), we only include ice as it is the only material with single phonons above $\sim$160 meV.
We note that the definition of $\bar{\sigma}_n \propto m_{\chi}^{-4}$ leads to the generally better reach at high masses\cite{Target_comparison_DM}.

\begin{figure*}[t]
\begin{center}
\includegraphics[trim = 14mm 55mm 25mm 59mm, clip, width=1\linewidth]{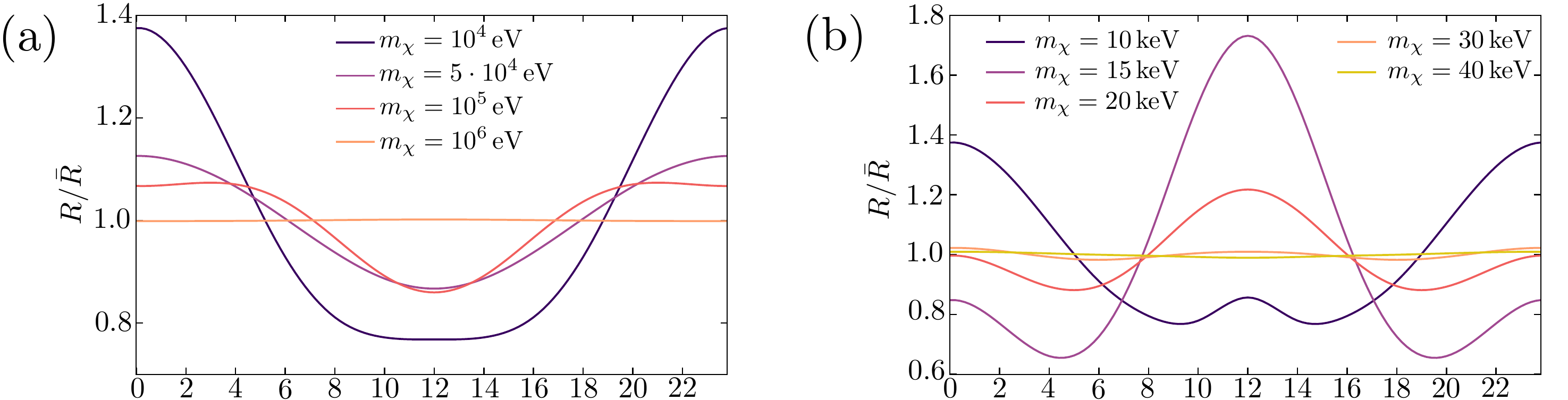}
\caption[]{Daily modulation for H$_2$O XI$_h$ ice with (a) a light dark photon and (b) a light scalar mediator for various $m_{\chi}$. }
\label{XIh_dailymod_both_med}
\end{center}
\end{figure*}

We finally calculate the daily modulation rate assuming that the z-axis of the target crystal is aligned with the DM wind at $t=0$. Each rate is normalised to the average rate $\bar{R}$ over one day. For a dark photon mediator we find a strong modulation for a mass range of $m_{\chi}=10^4-10^5$ eV (Fig.~ \ref{XIh_dailymod_both_med} (a)). We also find a significant daily modulation for the scalar mediator case (Fig.~\ref{XIh_dailymod_both_med} (b)), especially in the low 10 keV range.\\
It is important to note that high-energy optical phonons tend to undergo rapid downconversion to lower-energy modes. For instance, the reported smallest phonon linewidths for typical semiconductors such as GaAs are $\sim$1 - 2~cm$^{-1}$\cite{Yang_et_al:2020, Irmer_et_al:1996, Kang_et_al:2008}, which is approximately half of those calculated for ice (4 - 5~cm$^{-1}$) at 0~K\cite{anom_isotropic_properties_ice_phonon_lifetimes}. As a result, the range of phonon lifetimes (0.5 - 2.1\,ps) is correspondingly smaller than those observed in inorganic semiconductors (see Supplemental Materials), and can be attributed to the strongly anharmonic behavior of phonons in ice\cite{anom_isotropic_properties_ice_phonon_lifetimes}.\\
Our results suggest that ice H$_2$O is a competitive candidate target for single-phonon based light DM detection both in the low-mass and high-mass range (here we considered ice XI$_h$, but our conclusions are generally applicable to other ice polymorphs). Molecular units provide ultra-low-mass optical phonons (down to 6\,meV) giving excellent reach in the low-mass DM range for dark photon mediators. We expect similar reach to low masses for polar semiconductors containing heavy ions (as e.g. CsI) or molecular crystals where the relative effective masses of the ions/molecules result in low-frequency optical phonons.  For near-term experiments, we find that the broad range of single phonons in ice up to 400 meV gives it enhanced reach for higher experimental thresholds compared to alternative proposals such as multiphonon excitations. However, the lower density of ice due to its van-der-Waals bonding results in a reduced reach compared to other solid-state targets; this can be addressed by the growth of large single crystals of ice which is already well explored\cite{Khusnatdinov/Petrenko:1996}. Finally, while ice H$_2$O is cheap, earth-abundant, and has been extensively characterized, it remains an engineering challenge as to the best way to incorporate a material that is a liquid at room temperature into a (cryogenic) detector architecture.

We thank Dan Carney, Tanner Trickle, Kathryn Zurek, Katherine Inzani and Thomas Harrelson for helpful discussions. This work was supported
by the US Department of Energy under contract DEAC02-05CH11231 and Quantum Information Science Enabled Discovery (QuantISED) for High Energy Physics
grant KA2401032. Computational resources were provided by the National Energy Research Scientific Computing Center and the Molecular Foundry, DOE Office of Science User Facilities supported by the Office of Science, U.S. Department of Energy under Contract No. DEAC02-05CH11231. The work performed at the Molecular Foundry was supported by the Office of Science, Office of Basic Energy Sciences, of the U.S. Department of Energy under the same contract.

\bibliographystyle{apsrev4-1} 
\bibliography{references}

\end{document}